\documentclass[prd,twocolumn,groupedaddress,showpacs,nofootinbib]{revtex4}
\usepackage{graphicx}
\usepackage{dcolumn}
\usepackage{amssymb}
\usepackage{mathrsfs}
\usepackage{amsmath}
\usepackage{epsfig}
\usepackage[dvips]{color}
\usepackage{hhline}

\begin{document}

\title{High-scale leptogenesis with three-loop neutrino mass generation and dark matter}

\author{Pei-Hong Gu}

\email{peihong.gu@sjtu.edu.cn}

\affiliation{Department of Physics and Astronomy, Shanghai Jiao Tong University, 800 Dongchuan Road, Shanghai 200240, China}

\begin{abstract}

We demonstrate a common origin for high-scale leptogenesis and three-loop neutrino mass generation. Specifically we extend the standard model by two real singlet scalars, two singly charged scalars carrying different lepton numbers and two or more singlet fermions with Majorana masses. Our model respects a softly broken lepton number and an exactly conserved $Z_2^{}$ discrete symmetry. Through the lepton-number-violating decays of the real scalars and then the lepton-number-conserving decays of the charged scalars, we can obtain a lepton asymmetry stored in the standard model leptons. This lepton asymmetry can be partially converted to a baryon asymmetry by the sphaleron processes. The interactions for this leptogenesis can also result in a three-loop diagram to generate the neutrino masses. The lightest singlet fermion can keep stable to serve as a dark matter particle.

\end{abstract}


\maketitle

\section{Introduction}

The phenomena of neutrino oscillations have been established by the atmospheric, solar, accelerator and reactor neutrino experiments. This means the existence of three flavors of massive and mixing neutrinos which are beyond the standard model (SM) \cite{patrignani2016}. Meanwhile, the cosmological observations stringently constrain the neutrino masses should be in the sub-eV region \cite{patrignani2016}. Currently the most popular scheme for the neutrino mass generation is the famous seesaw \cite{minkowski1977,mw1980,flhj1989,ma1998} mechanism which can highly suppress the neutrino masses by a small ratio of the electroweak scale over a newly high scale. The seesaw scale can be lowed if we do some fine tuning on the related couplings. In the usual seesaw models, the lepton-number-violating interactions for the neutrino mass generation can also accommodate a leptogenesis \cite{fy1986,lpy1986,fps1995,ms1998,bcst1999,di2002,gnrrs2003,bbp2005,dnn2008} mechanism to generate the baryon asymmetry in the universe. In this scenario the seesaw and the leptogenesis are realized at a same scale.

Alternatively, some TeV-scale fields can help us to obtain the small neutrino masses at loop level \cite{zee1980,zee1985,bl2001,knt2003,cs2004,ma2006,kss2011,acmn2014,amn2014,ma2015,gms2016}. In this scenario, the neutrino masses may be suppressed by the charity besides the loop factors. For example, Krauss, Nasri and Trodden (KNT) ever proposed an interesting model with two TeV-scale singly charged scalars and one Majorana singlet fermion to give the neutrino masses at three-loop level \cite{knt2003}. The Majorana singlet fermion can keep stable to serve as a dark matter particle. In order to fulfill the neutrino oscillation data which require at least two nonzero neutrino mass eigenvalues, the KNT model should contain two or more Majorana singlet fermions \cite{cs2004}. Although the KNT model has an advantage of testability at colliers, it cannot explain the cosmic baryon asymmetry.

In this work we will slightly extend the KNT model by two real singlet scalars in order to demonstrate an interesting scenario that a high-scale leptogenesis can be consistent with a testable neutrino mass generation. Specifically, the real singlet scalars are very heavy so that their decays can be responsible for the leptogenesis. Meanwhile, we can obtain the KNT model by integrating out these real singlet scalars.

\section{The model}

We denote the non-SM fields by
\begin{eqnarray}
&&N_R^{}(1,1,0)(0)\,,~~\delta(1,1,+1)(-2)\,,~~\xi(1,1,+1)(-1)\,,\nonumber\\
[2mm]
&&\sigma(1,1,0)(0)\,.
\end{eqnarray}
Here and thereafter the first brackets following the fields describe the transformations under the $SU(3)_c^{}\times SU(2)_L^{}\times U(1)^{}_{Y}$ gauge groups, while the second brackets are the lepton numbers. We assume the lepton number can be softly broken. Furthermore, our model respects a $Z_2^{}$ discrete symmetry under which the fields transform as
\begin{eqnarray}
(\textrm{SM}\,,\,\delta)\stackrel{Z_2^{}}{\longrightarrow}(\textrm{SM}\,,\,\delta)\,,~(N_R^{}\,,\,\xi\,,\,\sigma) \stackrel{Z_2^{}}{\longrightarrow}-(N_R^{}\,,\,\xi\,,\,\sigma)\,.
\end{eqnarray}
The $Z_2^{}$ symmetry will not be broken at any scales. This means the real singlet scalars $\sigma$ will not be allowed to obtain any non-zero vacuum expectation values.

\begin{figure*}
\vspace{4cm} \epsfig{file=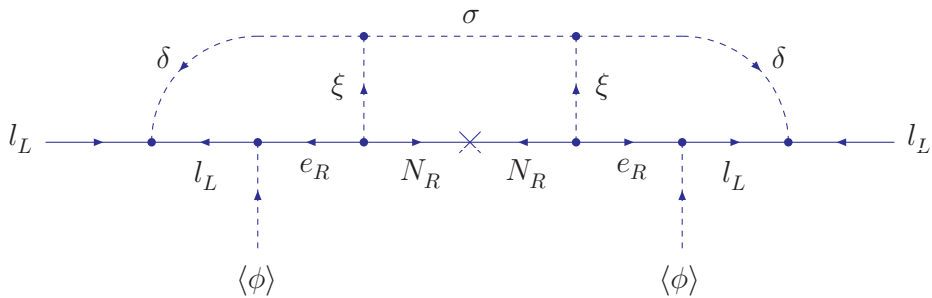, bbllx=5.25cm, bblly=6.0cm,
bburx=15.25cm, bbury=16cm, width=8cm, height=8cm, angle=0,
clip=0} \vspace{-8cm} \caption{\label{numass} The three-loop diagram for neutrino mass generation.}
\end{figure*}

Under the softly broken lepton number and the exactly conserved $Z_2^{}$ discrete symmetry, the Lagrangian should include
\begin{eqnarray}
\label{lag}
\mathcal{L}&\supset&-\frac{1}{2}M_{\sigma_{i}^{}}^2\sigma^2_{i}-(\mu_\delta^2+ \lambda^{}_{\delta\phi}\phi^\dagger_{}\phi)\delta^\dagger_{}\delta\nonumber\\
&&-(\mu_\xi^2+ \lambda^{}_{\xi\phi}\phi^\dagger_{}\phi)\xi^\dagger_{}\xi - \frac{1}{2}M_{N_i^{}}^{}(\bar{N}_{Ri}^{}N_{Ri}^c+\textrm{H.c.})\nonumber\\
&&-\left[\rho_{i}^{}\sigma^{}_i\xi^\dagger_{}\delta+\frac{1}{2}(f_{\delta}^{})_{\alpha\beta}^{}\delta\bar{l}_{L\alpha}^c i\tau_2^{}l_{L\beta}^{}+(f_\xi^{})_{\alpha i}^{} \xi^{} \bar{e}_{R\alpha}^c N_{Ri}^{}\right.\nonumber\\
&&\left.+\textrm{H.c.}]\right.-y_\alpha^{}(\bar{l}_{L\alpha}^{}\phi e_{R\alpha}^{}+\textrm{H.c.})\,.
\end{eqnarray}
Here $\phi$ denotes the SM Higgs scalar while $l_{L}^{}$ and $e_R^{}$ are the SM leptons,
\begin{eqnarray}
\phi(1,2,+\frac{1}{2})(0)&=&\left[\begin{array}{c} \phi^{+}_{} \\
[2mm] \phi^{0}_{}\end{array}\right]\,,\nonumber\\
[2mm]
l_{L\alpha}^{}(1,2,-\frac{1}{2})(+1)&=&\left[\begin{array}{c} \nu^{}_{L\alpha} \\
[2mm] e_{L\alpha}^{}\end{array}\right]\,,\nonumber\\
[2mm]
e_{R\alpha}^{}(1,1,-1)(+1)&&(\alpha=e\,,~\mu\,,~\tau)\,.
\end{eqnarray}
Obviously, the singlet fermions $N_{Ri}^{}$ can form the Majorana fermions as follows,
\begin{eqnarray}
N_i^{}=N_{Ri}^{}+N_{Ri}^c=N_i^c\,.
\end{eqnarray}

We emphasize that the cubic terms among the non-SM scalars, i.e. the $\rho_i^{}$-terms in Eq. (\ref{lag}), are the unique source for the lepton number violation. In addition, the two parameters $\rho_{1,2}^{}$ are always allowed to have a relative phase. As we will show later this phase provides the necessary CP violation for the leptogenesis.

\section{Neutrino masses and dark matter}

As shown in Fig. \ref{numass}, the non-SM scalars and fermions can mediate a three-loop diagram in association with the Yukawa couplings for generating the SM lepton masses. Clearly, after the electroweak symmetry breaking, this three-loop diagram will contribute a Majorana mass term of the left-handed neutrinos. Since the real singlet scalars $\sigma$ are very heavy, they can be integrated out from Eq. (\ref{lag}). The resulting Lagrangian then can contain a sizable quartic coupling between the singly charged scalars $\delta$ and $\xi$, i.e.
\begin{eqnarray}
\mathcal{L}\supset -\kappa(\delta^\dagger_{}\xi)^2_{}+\textrm{H.c.}~~\textrm{with}~~\kappa=\sum_i^{}\frac{\rho_i^2}{M_{\sigma_i^{}}^2}\,.
\end{eqnarray}
We hence obtain the KNT model where the lightest one of the Majorana fermions $N_i^{}$ can be a stable dark matter particle. For simplicity, we will not repeat the details of the neutrino masses and the dark matter \cite{knt2003,cs2004,bergstrom2012}.

\section{Leptogenesis}

\begin{figure*}
\vspace{5.5cm} \epsfig{file=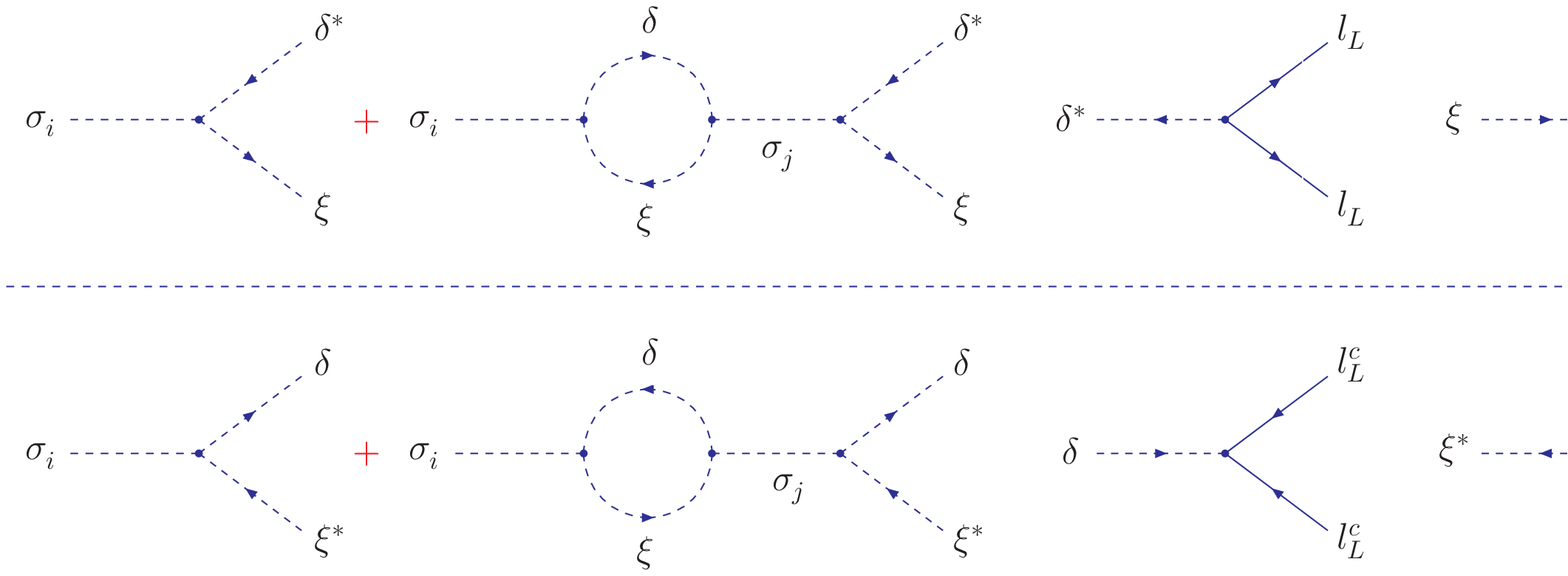, bbllx=7.5cm, bblly=6.0cm,
bburx=17.5cm, bbury=16cm, width=7cm, height=7cm, angle=0,
clip=0} \vspace{-6.75cm} \caption{\label{HSdecay} The real scalars $\sigma_i^{}$ decay into the charged scalars $\delta$ and $\xi$ which subsequently decay into the SM leptons $l_L^{}$ and $e_R^{}$ as well as the fermion singlets $N_R^{}$. }
\end{figure*}

Fig. \ref{HSdecay} shows the lepton-number-violating decays of the real singlet scalars $\sigma_i^{}$ as well as the lepton-number-conserving decays of the singly charged scalar pairs $(\delta\,,~\delta^\ast_{})$ and $(\xi\,,~\xi^\ast_{})$. The real scalar decays can generate a lepton asymmetry stored in the charged scalars. Through the charged scalar decays, the SM leptons then can acquire a lepton asymmetry, which participates in the sphaleron processes so that it can be partially converted to a baryon asymmetry. We calculate the width in the real scalar decays at tree level,
\begin{eqnarray}
\Gamma_{\sigma_i^{}}^{}=\Gamma(\sigma_i^{}\rightarrow \delta^\ast_{}+\xi)+\Gamma(\sigma_i^{}\rightarrow \delta+\xi^\ast_{})=\frac{1}{8\pi}\frac{|\rho_{i}^{}|^2_{}}{M_{\sigma_i^{}}^{}}\,,
\end{eqnarray}
and the CP asymmetry at one-loop order,
\begin{eqnarray}
\varepsilon_{\sigma_i^{}}^{}&=&\frac{\Gamma(\sigma_i^{}\rightarrow \delta^\ast_{}+\xi)-\Gamma(\sigma_i^{}\rightarrow \delta+\xi^\ast_{})}{\Gamma_{\sigma_i^{}}^{}}\nonumber\\
&=&\frac{1}{8\pi}\frac{\textrm{Im}(\rho_i^{\ast 2}\rho_j^{2})}
{|\rho_{i}^{}|^2_{}}\frac{1}{M_{\sigma_j^{}}^{2}-M_{\sigma_i^{}}^{2}}\nonumber\\
&=&\frac{\sin2\alpha_{ji}^{}}{8\pi}\frac{|\rho_j^{}|^2_{}}{M_{\sigma_j^{}}^{2}-M_{\sigma_i^{}}^{2}}~~\textrm{with}~~\alpha_{ji}^{}=\arg\left(\frac{\rho_j^{}}{\rho_i^{}}\right)\,.
\end{eqnarray}

As an example, we assume the real scalar $\sigma_{1}^{}$ much lighter than the other one $\sigma_{2}^{}$. The final baryon asymmetry then should mainly come from the $\sigma_{1}^{}$ decays. For a numerical estimation, we define
\begin{eqnarray}
\label{rwidth}
K&=&\frac{\Gamma_{\sigma_{1}^{}}^{}}{2H(T)}\left|_{T=M_{\sigma_{1}^{}}^{}}^{}\right.\,,
\end{eqnarray}
where $H(T)$ is the Hubble constant,
\begin{eqnarray}
H=\left(\frac{8\pi^{3}_{}g_{\ast}^{}}{90}\right)^{\frac{1}{2}}_{}
\frac{T^{2}_{}}{M_{\textrm{Pl}}^{}}\,,
\end{eqnarray}
with $g_{\ast}^{}$ being the relativistic degrees of freedom during the leptogenesis epoch. In the strong washout region where
\begin{eqnarray}
1\ll K\lesssim  10^6_{}\,,
\end{eqnarray}
the final baryon asymmetry can be simply described by \cite{kt1990}
\begin{eqnarray}
\eta_B^{}&=&\frac{n_B^{}}{s}\simeq -\frac{28}{79}\times \frac{\varepsilon_{\sigma_{1}^{}}^{}}{g_\ast^{}K z_f^{}}\nonumber\\
&&\textrm{with}~~z_f^{}=\frac{M_{\sigma_1^{}}^{}}{T_f^{}}\simeq 4.2(\ln K)^{0.6}_{}\,.
\end{eqnarray}
Here $n_B^{}$ and $s$, respectively, are the baryon number density and the entropy density, while the factor $-\frac{28}{79}$ is the sphaleron lepton-to-baryon coefficient. After fixing $g_\ast^{}=112.5$ (the SM fields plus two singly charged scalars as well as two singlet fermions) and inputting,
\begin{eqnarray}
\label{pchoice}
M_{\sigma_{1}^{}}^{}=|\rho_{1}^{}|=10^{14}_{}\,\textrm{GeV}\,,~~M_{\sigma_{2}^{}}^{}=|\rho_{2}^{}|=10^{15}_{}\,\textrm{GeV}\,,
\end{eqnarray}
we read
\begin{eqnarray}
&&K=137\,,~~z_f^{}=11\,,~~T_f^{}=9\times 10^{12}_{}\,\textrm{GeV}\,,\nonumber\\
&&\varepsilon_{\sigma_{1}^{}}^{}=5\times 10^{-5}_{}\left(\frac{\sin 2\alpha_{21}^{}}{1.25\times 10^{-3}_{}}\right)\,.
\end{eqnarray}
The baryon asymmetry then can arrive at an expected value,
\begin{eqnarray}
\label{basy}
\eta_B^{}= 10^{-10}_{}\left(\frac{\sin 2\alpha_{21}^{}}{1.25\times 10^{-3}_{}}\right)\,.
\end{eqnarray}

Ones may worry about the produced lepton asymmetry will be erased by some lepton-number-violating processes at low energies since Fig. \ref{numass} actually results in the dimension-5 Weinberg operators violating the lepton number by two units. Usually ones estimate these processes will decouple at a very high temperature \cite{fy1990},
\begin{eqnarray}
\label{estimation}
T=10^{12}_{}\,\textrm{GeV}\left[\frac{0.04\,\textrm{eV}^2_{}}{\textrm{Tr}(m_\nu^\dagger m_\nu^{})}\right]\,,
\end{eqnarray}
with $m_\nu^{}$ being the Majorana neutrino mass matrix. Therefore, no lepton asymmetry can survive above the temperature $T\sim 10^{12}_{}\,\textrm{GeV}$ if the neutrino masses arrives at an acceptable level. In our model, the effective dimension-5 operators are induced by integrating out the scalars $\sigma$, $\delta$ and $\xi$ as well as the fermions $N_R^{}$. However, the fields $\delta$, $\xi$ and $N_R^{}$ are near the TeV scale, i.e. their masses are lighter than the crucial temperature $T\sim 10^{12}_{}\,\textrm{GeV}$. So, the estimation (\ref{estimation}) is not consistent with the present scenario. Actually, in our model, the cubic terms among the scalars $\sigma$, $\delta$ and $\xi$ provide the unique source of the lepton number violation. After this lepton number violation is decoupled, no other lepton-number-violating processes can keep in equilibrium to wash out the produced lepton asymmetry.

\section{Summary}

In this paper we have simultaneously realized a high-scale leptogenesis and a low-scale neutrino mass generation. Specifically we have introduced two real singlet scalars to the KNT model which extended the SM by two singly charged scalars and two or more singlet fermions. The lepton-number-violating decays of the real scalars and then the lepton-number-conserving decays of the charged scalars can produce a lepton asymmetry stored in the SM leptons. This lepton asymmetry can be partially converted to a baryon asymmetry by the sphaleron processes. At the low energy scales, we can integrate out the real scalars to derive the KNT model, where the neutrino masses are induced at three-loop level while the dark matter particle is given by the lightest singlet fermion. For the variant KNT models \cite{acmn2014,amn2014}, we can consider two real triplet or quintuplet scalars.

\textbf{Acknowledgement}: The author was supported by the Recruitment Program for Young Professionals under Grant No. 15Z127060004, the Shanghai Jiao Tong University under Grant No. WF220407201 and the Shanghai Laboratory for Particle Physics and Cosmology under Grant No. 11DZ2260700. This work was also supported by the Key Laboratory for Particle Physics, Astrophysics and Cosmology, Ministry of Education.

\end{document}